\documentclass[aps,pra,showpacs,twocolumn,amssymb,superscriptaddress,floatfix]{revtex4-1}
\usepackage{bm}
\usepackage{color}
\usepackage{graphicx}
\usepackage{amsmath, amsthm}
\usepackage{everyshi}
\usepackage{hyperref}
\hypersetup{pdfstartview={FitH},pdfpagemode={UseNone},
            colorlinks,linkcolor=blue, citecolor=blue, urlcolor=blue,
            bookmarks=true, bookmarksopen=true, pdfnewwindow=true}

\graphicspath{{pict/}{}}

\newcommand\hrefBibPDF[3][]{}

\renewcommand{\Im}{\mathop{\mathrm{Im}}\nolimits}
\renewcommand{\Re}{\mathop{\mathrm{Re}}\nolimits}
\newcommand{\e}{{\rm e}}
\newcommand{\rmd}{{\rm d}}

\newcommand{\rmi}{{\rm i}}

\newcommand{\rot}{\mathop{\mathrm{rot}}\nolimits}

\begin{document}

%\title{Characterization of quantum nonlinear photonic circuits with laser light}
\title{Direct characterization of a nonlinear photonic circuit's wave function with laser light}
%Direct wave function characterization of photons generated in an arbitrary nonlinear quantum circuit
%\LARGE\textbf{Nonlinear biphoton generation tomography quantum-classical reciprocity on a chip}
%\Large\textbf{Nonlinear quantum-classical reciprocity for biphoton generation tomography on a chip}
%\LARGE\textbf{Characterization of an integrated multimode nonlinear device by classical sum-frequency generation}

\author{Francesco Lenzini}
\affiliation{Centre For Quantum Dynamics, Griffith University, Brisbane QLD 4111, Australia}

\author{Alexander N. Poddubny}
\affiliation{ITMO University, Saint Petersburg 197101, Russia}
\affiliation{Ioffe Institute, Saint Petersburg 194021, Russia}
\affiliation{Nonlinear Physics Centre, Research School of Physics and Engineering, Australian National University, Canberra ACT 2601, Australia}

\author{James Titchener}
\affiliation{Nonlinear Physics Centre, Research School of Physics and Engineering, Australian National University, Canberra ACT 2601, Australia}

\author{Paul Fisher}
\affiliation{Centre For Quantum Dynamics, Griffith University, Brisbane QLD 4111, Australia}

\author{Andreas Boes}
\affiliation{School of Engineering, RMIT University, Melbourne VIC 3000, Australia}

\author{Sachin Kasture}
\author{Ben Haylock}
\author{Matteo Villa}
\affiliation{Centre For Quantum Dynamics, Griffith University, Brisbane QLD 4111, Australia}

\author{Arnan Mitchell}
\affiliation{School of Engineering, RMIT University, Melbourne VIC 3000, Australia}

\author{Alexander S. Solntsev}
\author{Andrey A. Sukhorukov}
\affiliation{Nonlinear Physics Centre, Research School of Physics and Engineering, Australian National University, Canberra ACT 2601, Australia}

\author{Mirko Lobino}
\email{m.lobino@griffith.edu.au}
\affiliation{Centre For Quantum Dynamics, Griffith University, Brisbane QLD 4111, Australia}
\affiliation{Queensland Micro- and Nanotechnology Centre, Griffith University, Brisbane QLD 4111, Australia}

\begin{abstract}
Integrated photonics is a leading platform for quantum technologies including nonclassical state generation~\cite{Vergyris:2016-35975:SRP, Solntsev:2014-31007:PRX, Silverstone:2014-104:NPHOT, Solntsev:2016:RPH}, demonstration of quantum computational complexity~\cite{Lamitral_NJP2016} and secure quantum communications~\cite{Zhang:2014-130501:PRL}. As photonic circuits grow in complexity, full quantum tomography becomes impractical, and therefore an efficient method for their characterization~\cite{Lobino:2008-563:SCI, Rahimi-Keshari:2011-13006:NJP} is essential.
Here we propose and demonstrate a fast, reliable method for reconstructing the two-photon state produced by an arbitrary quadratically nonlinear optical circuit. By establishing a rigorous correspondence between the generated quantum state and classical sum-frequency generation measurements from laser light, we overcome the limitations of previous approaches for lossy multimode devices~\cite{Liscidini:2013-193602:PRL, Helt:2015-1460:OL}. We applied this protocol to a multi-channel nonlinear waveguide network, and measured a 99.28$\pm$0.31\% fidelity between classical and quantum characterization. This technique enables fast and precise evaluation of nonlinear quantum photonic networks, a crucial step towards complex, large-scale, device production.
\end{abstract}

\maketitle

\vspace{0.1cm}
Practical applications of quantum photonic technologies~\cite{OBrien:2009-687:NPHOT, Tanzilli:2012-115:LPR} require the integration of linear and nonlinear waveguides on a single device, where photons can be generated~\cite{Vergyris:2016-35975:SRP, Solntsev:2014-31007:PRX, Silverstone:2014-104:NPHOT, Solntsev:2016:RPH} and manipulated~\cite{Li_NJP2011}.
Spontaneous parametric down-conversion (SPDC) and spontaneous four-wave mixing are the two most common processes used for photon generation on chip with the former being the most efficient by far, needing only a few microwatts of pump power for generation rates exceeding several MHz~\cite{Tanzilli:2001-26:ELL, Zhang:2007-10288:OE}. Monolithic integration of SPDC sources with multi-port optical circuits has been achieved in several contexts, with applications in quantum communication~\cite{Martin:2012-25002:NJP}, quantum metrology~\cite{Vergyris:2016-35975:SRP}, spatial multiplexing of heralded single-photon sources~\cite{Meany:2014-L42:LPR}, quantum state generation in nonlinear waveguide arrays~\cite{Solntsev:2014-31007:PRX}, and small-scale demonstrations of reconfigurable quantum photonic circuits~\cite{Jin:2014-103601:PRL}. 

The near future of quantum photonics will involve an expansion in scale and applications of integrated circuits through large scale wafer fabrication. Successful fabrication procedures require effective device inspection techniques. However, the characterization of the two-photon state generated by a nonlinear waveguide network is a cumbersome experimental task~\cite{James:2001-52312:PRA}, requiring a quadratically increasing number of measurements and resources with system size. Here we propose and demonstrate a fast and practical method for the characterization of the two-photon wave function generated by an arbitrary waveguide device with quadratic nonlinearity that uses only laser probes and power measurements. This protocol has both fundamental and practical importance for the development of future integrated quantum photonics technologies since it could characterize waveguide networks with tens of modes in a fraction of a second when implemented with optimized hardware.

A method based on stimulated emission tomography (SET) was proposed~\cite{Liscidini:2013-193602:PRL} for predicting the two-photon wave function produced by a nonlinear device using the analogy between spontaneous nonlinear processes and their classical stimulated counterparts, i.e. difference-frequency generation or stimulated four-wave mixing. This technique was demonstrated for spectral characterization of two-photon states~\cite{Eckstein:2014-L76:LPR, Fang:2014-281:OPT,Jizan:2015-12557:SRP, Grassani:2016-23564:SRP}, and fast reconstruction of the density matrix of entangled-photon sources~\cite{Rozema:2015-430:OPT, Fang:2016-10013:OE}. 

However, SET has never been realized on multimode optical networks since it requires injection of the seed beam into the individual supermodes supported by the structure~\cite{Titchener:2015-33819:PRA}. A possible workaround is to inject the seed beam into each single channel individually then perform a transformation through supermode decomposition to obtain quantum predictions. Regardless, complete knowledge of the linear light dynamics inside the whole structure is required, making SET a multi-step procedure prone to errors and not applicable to ``black-box'' circuits. Additionally, SET is strictly valid only in the limit of zero propagation losses~\cite{Helt:2015-1460:OL}, posing a fundamental limitation for the characterization of real optical circuits. Characterization via sum-frequency generation (SFG), the reverse process of SPDC, gives exact results in the presence of any type of losses. However, so far it was formulated only for single, homogeneous waveguides~\cite{Helt:2015-1460:OL}, posing a stringent restriction for the characterization of more complex devices. 

\begin{figure}[t]
\centering
\includegraphics[width=\columnwidth]{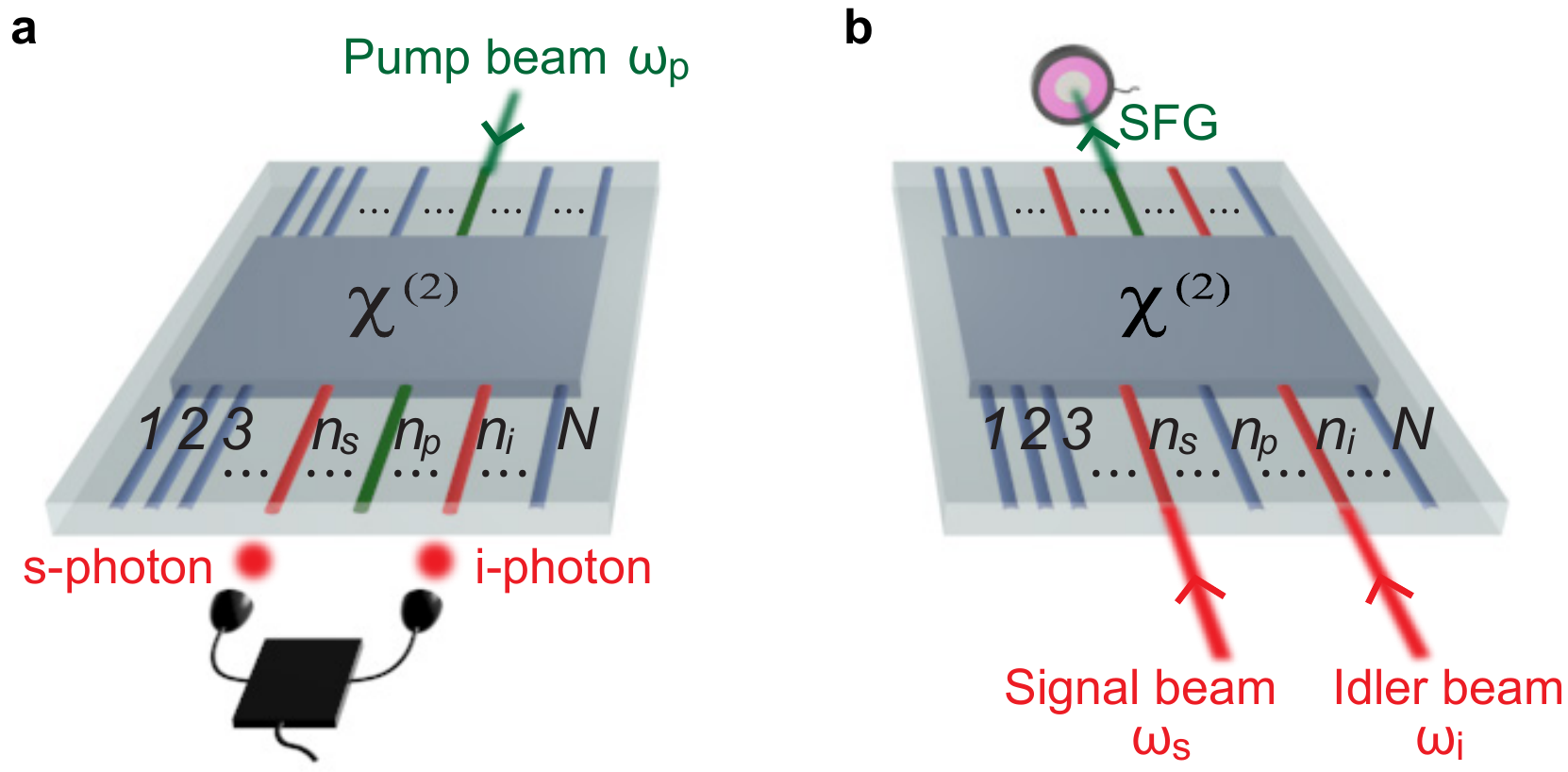}
\caption{\textbf{Scheme for the characterization of the biphoton state produced by an array of $N$ waveguides with an arbitrary $\chi^{(2)}$-nonlinear process.} \textbf{(a)}~SPDC: a pump beam is injected into waveguide $n_p$ at the input of the device. Photon-coincidence counting measurements between each pair of waveguides $(n_s,n_i)$ at the output are used to measure pair generation rates and relative absolute squared values of the wave function. \textbf{(b)}~SFG: Laser light at signal and idler frequencies is injected into waveguides $n_s$ and $n_i$ in the reverse direction of SPDC. Absolute photon-pair generation rates and relative absolute squared values of the wave functions can be predicted by direct optical power detection of the sum-frequency field emitted from waveguide $n_p$.}
\label{fig1}
\end{figure}

In this work, we uncover a fundamentally important equivalence between the biphoton wave function and the sum-frequency field generated by classical wave-mixing in the reverse direction of SPDC for any multimode nonlinear device, thus overcoming the limitations of previous approaches. Our theoretical analysis is based on the rigorous use of the Green-function method~\cite{Poddubny:2016-123901:PRL} (see Supplementary Information), and holds for arbitrarily complex second-order nonlinear circuits, in the presence of any type of losses. More importantly, the SFG-SPDC analogy can be expressed in any measurement basis, providing a simple experimental tool for the characterization of any ``black-box'' $\chi^{(2)}$-nonlinear process (see Fig.~\ref{fig1}).

%start of corrections introduced by ANP on 28 Nov 2016

Multimode SFG characterization can reconstruct any degree of freedom of the photonic state including spatial mode, frequency, time-bin, and polarization. Here, we illustrate its application in a ``black-box'' device with $N$ spatial modes of the same polarization, as schematically depicted in Fig.~\ref{fig1}. When a pump beam with frequency $\omega_p$ is injected into waveguide $n_p$ at the input of the device it produces, by SPDC, the biphoton state (see Fig.~\ref{fig1}a)
\begin{multline}
|\Psi_{\rm pair}\rangle=  \int\limits_{0}^{\infty}\!\!\int\limits_0^{\infty} \mbox{d} \omega_s \mbox{d} \omega_i \sum_{n_s , n_i=1}^N \Psi_{n_s  n_i}^{n_p} (\omega_s, \omega_i) \\\times\hat{a}^{\dagger}_{n_s} (\omega_s) \hat{a}^{\dagger}_{n_i} (\omega_i) |0\rangle \ ,
\end{multline}
where $n_s$ and $n_i$ are the indices for signal-idler output waveguide numbers, and $\hat{a}^{\dagger}_{n_s} (\omega_s), \hat{a}^{\dagger}_{n_i} (\omega_i)$ are the photon creation operators  in the waveguide $n_s$ with the frequency $\omega_s$ and in the waveguide $n_i$ with the frequency $\omega_i$, respectively. In the classical SFG process shown in Fig.~\ref{fig1}b, we use the reverse configuration to SPDC. Two beams  with signal frequency $\omega_s$ and power $P_s$ and idler  frequency $\omega_i=\omega_p-\omega_s$ and power $P_i$ are injected into the waveguides $n_s$ and $n_i$ from the SPDC output directions. The generated sum-frequency electric field $E_{n_s n_i}^{n_p}$ is detected from waveguide $n_p$.  

The sum-frequency field in the undepleted pump regime is directly proportional to the two-photon wave function  $\Psi_{n_s n_i}^{n_p} (\omega_s,\omega_i)$ (see Supplementary Information for the full derivaion).  We can use this correspondence to infer the squared amplitudes of the wave function elements by direct optical measurements of the  sum-frequency field power $P_{\rm SFG}$, and predict the absolute photon-pair generation rates for SPDC through the relation:
\begin{equation}
\frac{1}{P_p} \frac{{\rm d}N_{\rm pair}}{{\rm d} \omega_s {\rm d}t} =  \frac{\omega_i \omega_s}{2\pi\omega_p^2} \eta^{\rm SFG}_{n_s n_i} (\omega_s,\omega_i)\:.
\label{abs_rate}
\end{equation}
Here, $P_p$ is the power of the pump beam during SPDC, ${\rm d}N_{\rm pair}/{\rm d} \omega_s {\rm d}t$ is the rate of photon-pair coincidence counts per unit signal frequency, and $\eta^{\rm SFG}_{n_s n_i}\equiv P_{\rm SFG}/(P_{s}P_{i})$ is the  sum-frequency conversion efficiency. In addition to  the absolute values of the two-photon wave function intensity given by  Eq.~\eqref{abs_rate} and Eq.~(S24) in Supplementary, we are able to characterize  the relative phases of the  wave function components  by classical interferometric measurements of the generated sum-frequency field. Full spectral characterization of the biphoton state is obtained by repeating the procedure for different wavelengths of signal and idler beams, with an accuracy that is  limited only by the spectral resolution of the laser source.

\begin{figure*}[t!]
\centering
\includegraphics[width=\textwidth]{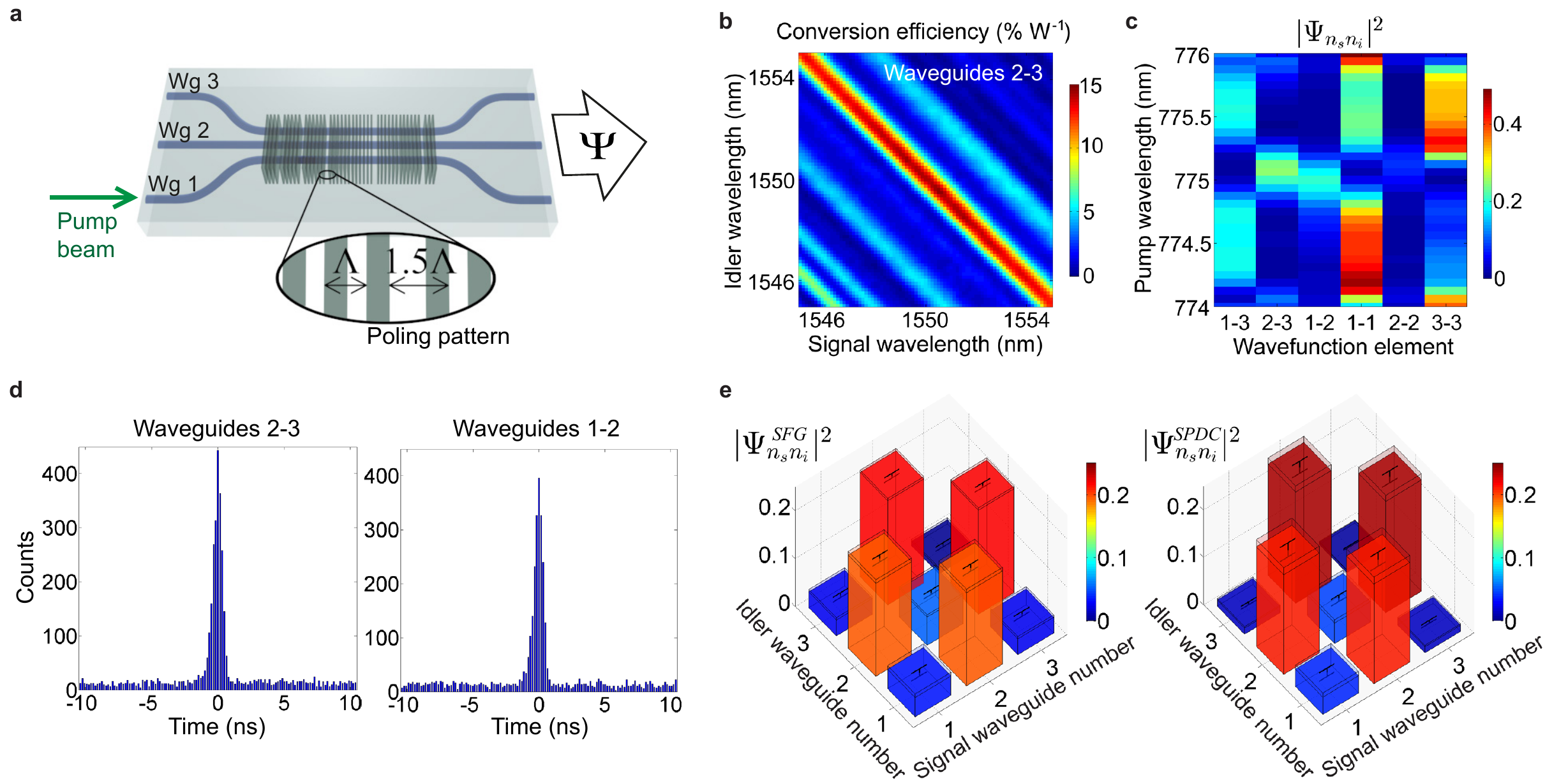}
\caption{\textbf{Comparison between SFG and SPDC measurements.} \textbf{(a)}~Schematic of the device used for biphoton state generation. The device is made of three coupled waveguides with specially introduced defects in the periodic poling pattern (inset). \textbf{(b)}~Measured classical sum-frequency conversion efficiency from waveguide 1 as a function of signal and idler wavelengths coupled to waveguides 2 and 3. \textbf{(c)}~Predicted squared relative amplitudes of the biphoton wave function, proportional to the SFG signal for different combinations of signal and idler in coupled waveguides vs. the pump wavelength in the degenerate regime ($\lambda_s=\lambda_i = 2 \lambda_p$). \textbf{(d)}~Time histogram for the photon coincidences between waveguides 2-3 and waveguides 1-2 for a 28.57 s acquisition time, a pump wavelength $\lambda_p=775 \ \mbox{nm}$, and a pump power $P_p=32 \pm 5 \ \mu \mbox{W}$. Time bin width is 82~ps. Complete data sets are in Supplementary Fig.~2. \textbf{(e)}~Normalized biphoton wave functions predicted by SFG (left) and measured by SPDC (right) for $\lambda_p=775 \ \mbox{nm}$.
%Error bars for SFG take into account the uncertainties in optical power measurements, and for SPDC account for the poissonian statistics of the detection process and uncertainties in the measured transmission of the different paths.
}
\label{fig2}
\end{figure*}

The SFG protocol was experimentally verified on an array of three evanescently coupled nonlinear waveguides schematically depicted in Fig.~\ref{fig2}a. The device was fabricated on a $Z$-cut lithium niobate substrate by the reverse proton exchange technique~\cite{Lenzini:2015-1748:OE, Korkishko:1998-1838:JOSA} and heated to $T= 84 \ ^\circ\mbox{C}$ to obtain phase matching centred at $\lambda$ = 1550~nm (see Methods). The three waveguides have an inhomogeneous and asymmetric poling pattern along the propagation direction, with five defects introduced by translating the poled domains by half a poling period $\Lambda$ at different locations of the array (see Methods). This design is based on the recently developed concept for quantum state engineering with specialized poling patterns~\cite{Titchener:2015-33819:PRA}. 

We performed the SFG measurements by coupling two frequency tunable lasers into the device and measuring sum-frequency generation from waveguide 1. Figure~\ref{fig2}b shows the SFG efficiency $\eta_{SFG}$ as a function of signal and idler wavelengths coupled to the waveguides 2 and 3, respectively. Similar data were taken for all input combinations (see Supplementary Fig.~1). A maximum normalized conversion efficiency of $\simeq 15 \ \% ~ \mbox{W}^{-1}$ was measured for a sum-frequency wavelength $(\lambda_s^{-1}+\lambda_i^{-1})^{-1}=775 \ \mbox{nm}$. 

Figure~\ref{fig2}c shows the squared biphoton wave function elements $|\Psi_{n_s n_i}^{SFG}|^2$ predicted from the SFG measurements as a function of the pump wavelength, in the degenerate regime, when $\lambda_s = \lambda_i = 2 \lambda_p$, showing a strong dependence of the generated state on the pump wavelength. Importantly, such classical data measurements are essentially instantaneous.

To verify the validity of our approach we directly measured the biphoton state generated by the device by coupling a $\lambda_p=775 \ \mbox{nm}$ pump laser into waveguide 1 and measuring the down-converted photon pairs from all different output combinations (see Supplementary Fig.~2). The pump laser was coupled in the opposite direction with respect to the SFG lasers. Photon pairs were filtered with a 6~nm band-pass filter centred at $\lambda_c=1550 \ \mbox{nm}$ to restrict the SPDC emission bandwidth to the range measured by SFG. Figure~\ref{fig2}d shows two characteristic time histograms of photon coincidences for waveguides~2-3 and waveguides~1-2 outputs acquired by two avalanche photodiodes and a time tagging module. Coincidence-to-accidental-ratio~(CAR) is $\simeq 24.5$.

The squared amplitudes of the normalized wave function elements predicted by SFG and those directly measured from SPDC are shown in Fig.~\ref{fig2}e (see Methods for details on calculations, and Supplementary Table 1). The two matrices have a fidelity $F=\sum_{n_s n_i} \sqrt{|\Psi_{n_s n_i}^{SFG}|^2 |\Psi_{n_s n_i}^{SPDC}|^2}= 99.28 \pm 0.31 \ \%$. The elements $|\Psi_{n_s n_i}^{SFG}|^2$ were obtained by integrating the data in Fig.~\ref{fig2}b corresponding to a pump wavelength $\lambda_p=775 \ \mbox{nm}$ over a bandwidth of 6~nm. 

\begin{figure*}[t]
\centering
\includegraphics[width=\textwidth]{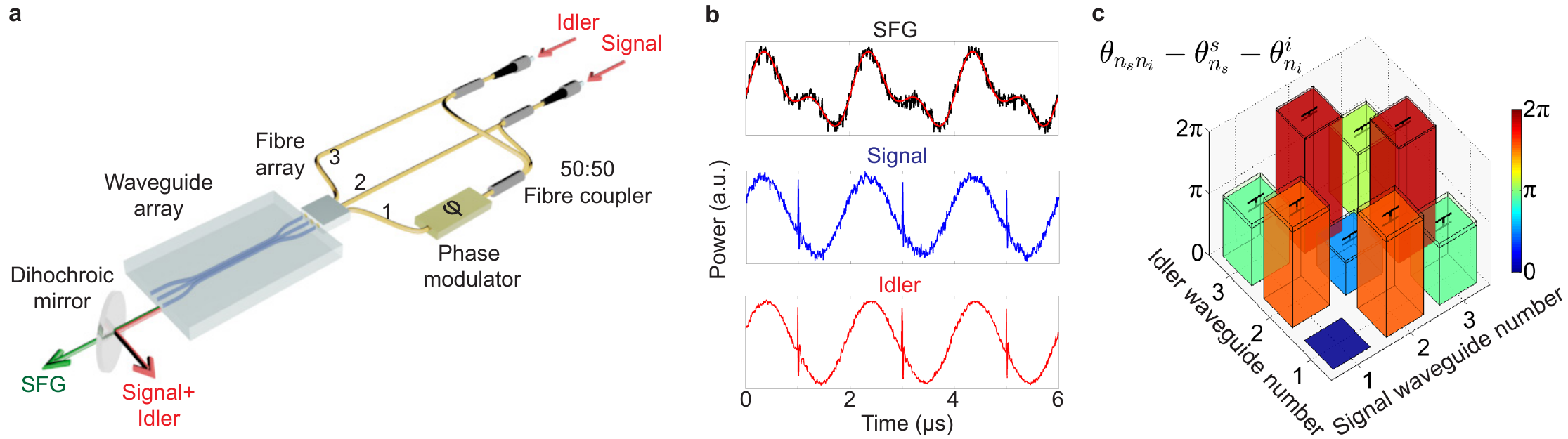}
\caption{\textbf{Measurement of the relative phases between wave function elements by SFG.} \textbf{(a)}~Schematic of the experimental setup for input into waveguides 2 and 3. Signal and idler beams are split and recombined with a network of 50:50 fibre couplers and injected into the three waveguides with a fibre V-groove array. An electro-optic phase modulator is used to generate an interference pattern between the sum-frequency fields generated from the combinations of signal and idler beams in waveguides 2-3 and waveguides 1-1. SFG and signal-idler beams are collected in free-space at the output of waveguide~1 with a lens of 0.5 NA (not shown in the figure) and separated with a dichroic mirror. A wavelength-division multiplexer (not shown in the figure) is used to separate signal and idler wavelengths. \textbf{(b)} Oscilloscope traces obtained by collecting the beams with three different photodiodes for a modulation frequency $f=500 \ \mbox{KHz}$. The three traces are used to measure the relative phase between the wave function elements $\Psi_{23}$ and $\Psi_{11}$. Solid red line is the theoretical fit (see Methods for details). \textbf{(c)}~ Relative phases between wave function elements measured for all the combinations of signal-idler beams in the three waveguides. Waveguide~1 is the fixed reference for all the phase measurements. Measurements are performed for a signal wavelength $\lambda_s=1550.12 \ \mbox{nm} $ and an idler wavelength $\lambda_i=1556.55 \ \mbox{nm}$. The sample was heated up to $T= 108$~$^{\circ}$C to get a phase matching condition centered at $2(\lambda_s^{-1}+\lambda_i^{-1})^{-1} \simeq 1553.3 \mbox{nm}$. See Methods for a calculation of the error bars.}
\label{fig3}
\end{figure*}

From Eq.~(\ref{abs_rate}), using the SFG measurements, we calculated a photon pair generation rate $N_{SFG} \simeq 2.4~\mbox{MHz}$, which is the sum of the rates from all 6 output combinations. Direct measurement of this rate from SPDC data gives $N_{SPDC} \simeq 1.7~\mbox{MHz}$, showing a good qualitative agreement between the two values. Overestimation of the detector efficiencies~($\eta_1=8\%$, $\eta_2=10\%$ as provided by the manufacturer)  is the likely cause of the discrepancy.

Importantly, our method also allows direct characterization of the phases between the wavefunction elements, by performing interferometric detection of the generated sum-frequency. In our case, verification of the generated state by quantum state tomography would be experimentally difficult due to phase fluctuations between the different paths introduced by thermal and mechanical instabilities and the long acquisition times needed for photon coincidences counting. Hence, the SFG-phase characterization is only presented as a proof-of-concept and not directly verified by SPDC measurements. 

The experimental setup for phase measurements is illustrated in Fig.~\ref{fig3}a for input into waveguides 2 and 3. The depicted procedure allows us to infer the relative phases between wave function elements $\theta_{n_s n_i}$ up to the phases of signal and idler beams $-(\theta_{n_s}^s + \theta_{n_i}^i)$ measured at the output of waveguide~1 (see Fig.~\ref{fig3}b, Methods and Supplementary Fig.~3.). The predicted phase structure of the wave function is shown in Fig.~\ref{fig3}c. We note that the unknown phase multiplier $\exp[-\mbox{i}(\theta_{n_s}^s + \theta_{n_i}^i)]$ doesn't alter the degree of entanglement of the biphoton state. Thus, the set of measured phases can be directly used to infer the non-classical properties of the generated state such as the degree of entanglement through Schmidt decomposition~\cite{Nielsen:2011:QuantumComputation}. Based on the SFG phase and intensity measurements, we obtain a Schmidt number $S=1.59$ (see Methods for details on the calculation), which precisely characterizes the degree of spatial entanglement and cannot be obtained with only photon correlations.

The SFG characterization method proposed here provides a fast and resource efficient path for the practical characterization and development of monolithically integrated networks based on a rigorous theoretical proof of quantum-classical correspondence. This technique can be applied to any arbitrary ``black-box'' second-order nonlinear device and supports the development of integrated photon sources and large-scale quantum photonics technologies.

%\vspace{0.2cm}
\section*{Acknowledgements}
%\footnotesize

This work was supported by the Australian Research Council (ARC) under the Grants DP140100808 and DP160100619, the Centre of Excellence for Ultrahigh bandwidth Devices for Optical Systems (CUDOS), and the Griffith University Research Infrastructure Program. BH and PF are supported by the Australian Government Research Training Program Scholarship. This work was performed in part at the Queensland node of the Australian National Fabrication Facility, a company established under the National Collaborative Research Infrastructure Strategy to provide nano-and microfabrication facilities for Australia's researchers.

%\vspace{0.2cm}
\section*{Author contributions}
%\footnotesize

A.N.P., J.T., A.S.S., and A.A.S. developed the theory for the SFG-SPDC analogy.
F.L., A.B., S.K., and B.H. fabricated the nonlinear device.
F.L., P.F., A.B., and M.V. performed the experimental measurements.
A.M., A.S.S., A.A.S., and M.L. supervised the project.
F.L. wrote the manuscript with contributions from all authors.

\newpage
%\clearpage
%\newpage
\bibliographystyle{ref_links}
\bibliography{db_sfg_paper_lenzini}
%\bibliography{Lenzini_SFG_bibliography}
%\bibliographystyle{naturetest}

%\footnotesize

%\newpage
%\clearpage
%\newpage

%\Large
%\textbf{Methods}
%\small

\section*{Methods}

%\vspace{0.2cm}

\textbf{Fabrication of the waveguide array.} The waveguides were fabricated on a Z-cut lithium niobate wafer via reverse-proton-exchange. A titanium mask was used to pattern the channels with a width of $8.5 \ \mu \mbox{m}$ with a coupling region of $2.96 \ \mbox{cm}$ in length and a distance between waveguide centres of $11.6 \ \mu \mbox{m}$.
Proton exchange was performed by immersing the sample in a hot benzoic acid bath creating a $1.85 \ \mu \mbox{m}$ doped layer. Subsequent annealing in air for 7 hours at $328 \ ^{\circ} \mbox{C}$ and reverse proton exchange for 8.5 hours at the same temperature were performed. 

The poling area is 2 cm long and centered in the middle of the array. The poling pattern was generated by standard electric-field poling and has a poling period $\Lambda=16.07 \ \mu \mbox{m}$ and a 50:50 duty cycle. Defects in the poling pattern are located at 1173.11~$\mu$m, 4017.50~$\mu$m, 7536.83~$\mu$m, 13723.78~$\mu$m and 19042.95~$\mu$m from the beginning of the poling region. S-Bends with a sinusoidal shape and a 5.5 mm length were used at the input and the output of the array to achieve a $127 \ \mu \mbox{m}$ separation between waveguide centres matching the pitch of standard fibre V-groove arrays. The input facet of the chip is polished at an $8^{\circ}$ angle to avoid back-reflections into the waveguides.

\textbf{Setup for SFG measurement.} Signal and idler beams, generated by two tunable laser diodes with 100~KHz linewidth, were injected into each pair of waveguides with a fibre V-groove array. All the beams were collected in free-space at the output of the waveguides with a lens with 0.5 NA. SFG and signal-idler wavelengths were separated with a dichroic mirror. SFG power from the output of waveguide~1 and signal-idler powers from the outputs of all three waveguides were then measured with two standard power meters. The measured powers were corrected for Fresnel losses at the chip interface and used to calculate the normalized SFG conversion efficiency at the output of the array.
SFG conversion efficiencies for the single channel inputs were measured by combining signal and idler beams with a 50:50 fibre coupler. The measurement process was automated with Labview.

\textbf{Experimental setup for SPDC measurements.} A pump beam with $775 \ \mbox{nm}$ wavelength and 100~KHz linewidth was generated by second-harmonic generation in a periodically poled lithium niobate waveguide and injected into waveguide~1 with a lens of 0.5 NA. The three outputs were collected with a fibre V-groove array, and photon coincidences between each pair of waveguides were measured with two gated InGaAs avalanche photodiodes and a time-tagging module.
A filtering stage in free-space, made from a set of 5 long-pass filters and a band-pass filter, was used to attenuate the pump beam by 150 dB and to narrow down
the measured SPDC bandwidth to 6~nm. Photon coincidences from the single channels were measured by splitting signal-idler photons with a 50:50 fibre coupler.

\textbf{Absolute photon pair generation rates and relative squared amplitudes of the wave function.} For each pair of waveguides $n_s,n_i$, the signal wavelength was scanned in steps of $\Delta\lambda=0.25 \ \mbox{nm}$ in a 6 nm bandwidth centered around 1550 nm. At each step $j$ the idler wavelength was set to $(\lambda_i)_j=(\lambda_p^{-1} - (\lambda_s)_j^{-1})^{-1}$, where $\lambda_p=775 \ \mbox{nm}$ is the pump wavelength for SPDC. Absolute photon pair generation rates were calculated by discretization of Eq.~(\ref{abs_rate}) through the relation
$$
\frac{1}{P_p} \frac{\mbox{d}N_{pair}}{\mbox{d}t} = \sum_j \eta^{SFG}_j \frac{\lambda_p^2}{(\lambda_s)_j (\lambda_i)_j}\frac{c \Delta \lambda}{ [(\lambda_s)_j]^2} \ ,
$$
where $\eta^{SFG}_j$ is the normalized sum-frequency conversion efficiency measured at each step $j$.
The pump power $P_p$ was measured during the SPDC characterization from the first output of the fibre array. %This value is corrected with the measured coupling efficiency of the $775 \ \mbox{nm}$ beam to a single mode fibre at $1550 \ \mbox{nm}$.
Relative squared amplitudes of the wave function elements were calculated as
$$
\left|\Psi_{n_s n_i}^{SFG}\right|^2=\frac{\left(\sum_j \eta^{SFG}_j\right)_{n_s n_i}}{\sum_{n_s,n_i}\left( \sum_j \eta^{SFG}_j\right)_{n_s n_i}} \ .
$$

\textbf{Photon pair generation rates and wave function square moduli from SPDC measurements.}
Absolute photon pair generation rates were calculated for each input combination $n_s,n_i$ as
$$
\frac{\mbox{d}N_{pair}}{\mbox{d}t} = \frac{C_{n_sn_i}}{ \Delta T \mu_{n_s} \mu_{n_i} \eta_1 \eta_2} \ ,
$$
where $\Delta T$ is the acquisition time, $\mu_{n_s}$ and $\mu_{n_i}$ are the total transmissions of the single channels (including paths from waveguides to detectors), and $\eta_1$, $\eta_2$ are the quantum efficiencies of the two detectors. $C_{n_s n_i}$ is the total number of counts in a time window equal to 2 FWHM centered around the coincidence peak acquired with the time tagging module. Accidental counts were measured in an equal time window away from the peak and subtracted from this value. 

To take into account fluctuations in the pump power ($\simeq 15 \ \%$) during the acquisition time, coincidence values were rescaled by using the average total number of counts measured from waveguide~2 as a common reference for the average power. Squared relative amplitudes of the wave function elements were calculated as

$$
\left|\Psi_{n_s n_i}^{SPDC}\right|^2=\frac{C_{n_s n_i}/(\mu_{n_s} \mu_{n_i})}{\sum_{n_s,n_i}\left[C_{n_s n_i}/(\mu_{n_s} \mu_{n_i})\right]} \ .
$$

\textbf{Error in the fidelity between correlation matrices.}
The error in the fidelity between the correlation matrices predicted by SFG and measured by SPDC was calculated with an iterative numerical algorithm with $N=10^{6}$ cycles. At each step we assigned to the two correlation matrices a random value calculated from a normal distribution with a sigma given by the error in the measurements. Average value and error in the fidelity were finally calculated from the simulated distribution.

\textbf{Second-harmonic generation contributions in SFG measurements}.
For SFG-power measurements second-harmonic generation~(SHG) contributions were first measured by inputting signal and idler beams into each channel individually. SHG powers were then subtracted from SFG-power measurements. The procedure was repeated and automated with Labview. For SFG-phase measurements, SFG and SHG contributions were separated at the output of the array with the aid of a diffraction grating.

\textbf{SFG-phase measurements.}
We describe, as an example, the procedure used for phase measurements for the case of waveguides~2-3. An equivalent procedure was used for all the other waveguides combinations. Let us call, with reference to Fig.~3a, $\Delta \phi_s$~($\Delta \phi_i$) the phase difference at the input of the array between signal(idler) beam injected into waveguide~2(3) and signal(idler) beam injected into waveguide~1(1) when no voltage is applied to the phase modulator. Thermal and mechanical fluctuations are negligible in the given acquisition time and the two phase differences are assumed to be constant. The goal of the characterization is to measure the value $\theta^{SFG} - \theta^{s}- \theta^{i}$. $\theta^{SFG}$ is the relative phase between the sum-frequency fields generated from the combinations of signal-idler beams injected into waveguides~2-3 and signal-idler beams injected into waveguides~1-1 introduced by the array at the output of waveguide~1. $\theta^s$~($\theta^i$) is the phase difference between the signal(idler) beam injected into waveguide~2(3) and the signal(idler) beam injected into waveguide~1(1) introduced by the array at the output of the same waveguide. When a phase modulation with frequency $f$ is applied to the first channel at the input, the three generated interference patterns can be expressed as
\begin{flalign}
&\lvert E^{SFG}_{23}+E^{SFG}_{11}\rvert^2 \propto \cos(\theta^{SFG}+\Delta \phi_s + \Delta \phi_i - 4\pi f t)\nonumber\\ 
				&	+\mbox{constant terms},\nonumber\\
&\lvert E^{s}_{2}+E^{s}_{1}\rvert^2 \propto \cos(\theta^{s}+\Delta \phi_s - 2\pi f t) + \mbox{constant terms}, \nonumber\\ 
&\lvert E^{i}_{3}+E^{i}_{1}\rvert^2 \propto \cos(\theta^{i}+ \Delta \phi_i - 2\pi f t) +  \mbox{constant terms}. \nonumber 
\end{flalign}

To obtain the desired values, the acquired oscilloscope traces for signal and idler beams are fitted with the two functions $y_s= a_s \cos(c_s - 2\pi f t) +d_s$ and $y_i= a_i \cos(c_i - 2\pi f t)+d_i$. A fast fourier transform analysis of the oscilloscope trace for SFG reveals that the signal is made of a fast component oscillating with frequency $2f$ and a slow component with frequency $f$ due to the interference of sum-frequency fields generated from combinations other than waveguides~2-3 and waveguides~1-1 (namely, waveguides~1-2 and waveguides~1-3). Hence, the SFG trace is fitted with the function  $y_{SFG}= a_1 \cos(c_1 - 2\pi f t) + a_2 \cos(c_2 - 4\pi f t) +d$. The desired phases are finally calculated as $\theta^{SFG} - \theta^{s}- \theta^{i}=c_2-c_s-c_i$. Uncertainties in the measurements are obtained from the confidence bounds in the least squares fitting procedure.

\textbf{Calculation of the Schmidt number from SFG-phase measurements.} The degree of entanglement in the predicted state is calculated in a non-degenerate case for two fixed frequencies $\omega_s,\omega_i$ by expressing the biphoton state as

$$
|\Psi \rangle _{pair} = \sum_{n_s,n_i} \Psi_{n_s  n_i} \mathbf{a}^{\dagger}_{n_s} (\omega_s) \mathbf{a}^{\dagger}_{n_i} (\omega_i) |0\rangle \ .
$$

The amplitudes of the wave function elements $\Psi_{n_s  n_i}$ are calculated from the results of SFG-power measurements, while the relative phases are calculated from the results of SFG-phase measurements for the two fixed wavelengths used in the characterization. The Schmidt number $S$ is obtained trough the Schmidt decomposition for a bipartite system

$$
\Psi_{n_s n_i} = \sum_{j} \sqrt{S_j} U_{j n_s} V_{j n_i} \ ,
$$

as

$$
S=\frac{1}{\sum_j S_j^2} \ .
$$
\newpage
\begin{widetext}
\section*{Supplementary Information}
\textbf{Squared relative amplitudes of the wavefunction elements.}
\begin{table}[h!]
\small
\centering
\begin{tabular}{|c|c c|}
 \hline
 WF element & SPDC & SFG\\
 \hline\hline
 $|\psi_{13}|^2,|\psi_{31}|^2$ & $0.013 \pm 0.003$ & $0.033 \pm 0.005$ \\
 \hline
 $|\psi_{23}|^2,|\psi_{32}|^2$ & $0.23 \pm 0.01$ & $0.22 \pm 0.005$\\
 \hline
 $|\psi_{12}|^2,|\psi_{21}|^2$ & $0.212 \pm 0.015$ & $0.195 \pm 0.007$ \\
 \hline
 $|\psi_{11}|^2$ & $0.04 \pm 0.01$ & $0.04 \pm 0.008$ \\
 \hline
 $|\psi_{22}|^2$ & $0.046 \pm 0.007$ & $0.059 \pm 0.005$ \\
 \hline
 $|\psi_{33}|^2$ & $0.006 \pm 0.002$ & $0.009 \pm 0.004$ \\
 \hline
 \end{tabular}
\caption{\textbf{Squared relative amplitudes of the wavefunction elements.} Errors for SPDC take into account the poissonian statistics of the detection process and uncertainties in transmission measurements. Errors for SFG take into account uncertainty in optical power measurements. The fidelity between the two matrices is ($99.28 \pm 0.31$)\%.}
\end{table}
%%%%%%%%%%%%%%%%%%%%%%%%%%%%%%%%%%%%%%
\setcounter{equation}{0}
\renewcommand{\theequation}{S\arabic{equation}}
\renewcommand{\thefigure}{S\arabic{figure}}    
\setcounter{figure}{0}
%%%%%%%%%%%%%%%%%%%%%%%%%%%%%%%%%%%%%%
\\

\textbf{SPDC-SFG correspondence for a general ``black-box'' $\chi^{(2)}$-nonlinear process.~}

%%%%%%%%%%%%%%%%%%%%%%%%%%%%%%%%%%%%%%
Here we present  a derivation of the correspondence between SPDC process and the SFG process in the reversed geometry  for an arbitrary  $\chi^{(2)}$-nonlinear structure, that is reciprocal in the linear regime. This proof generalizes the Lorentz reciprocity theorem \cite{BornWolf:1999} in the form
\begin{equation}
\int\rmd^{3}r\bm P_{1}(\bm r)\cdot \bm E_{2}(\bm r)=
\int\rmd^{3}r\bm P_{2}(\bm r)\cdot \bm  E_{1}(\bm r)\label{eq:Lorentz}
\end{equation}
that links the electric field distribution  $\bm E_{1}(\bm r)$ and $\bm E_{2}(\bm r)$, induced by the polarization distributions $\bm P_{1}$ and $
\bm P_{2}$, respectively.

The biphoton wavefunction  in the SPDC regime \cite{Poddubny:2016-123901:PRL} reads
\begin{equation}\label{eq:Tsi1}
\Psi(\bm r_{s},\bm r_{i},\sigma_{s},\sigma_{i},\omega_{s},\omega_{i})=\int\rmd^{3} r_{0}
G_{\sigma_{s}\alpha}(\bm r_{s},\bm r_{0};\omega_{s})
G_{\sigma_{i}\beta}(\bm r_{i},\bm r_{0};\omega_{i})\chi^{{(2)}}_{\alpha\beta\gamma}E_{p,\gamma}(\bm r_{0})\:,
\end{equation}
where $G$ is the electromagnetic tensor Green function satisfying the equation
\begin{equation}\label{eq:Gtilde2}
 [\rot\rot-\left(\frac{\omega}{c}\right)^2\varepsilon(\bm r)]\hat G(\bm r,\bm r';\omega)=4\pi \left(\frac{\omega}{c}\right)^2 \delta(\bm r-\bm r')\:,
\end{equation}
$\bm E_{p}$ is the pumping wave and $\chi^{{(2)}}$ is the nonlinear susceptibility tensor. The indices $\sigma_{i}$ and $\sigma_{s}$ label signal and idler polarizations, respectively. 

On the other hand, the nonlinear wave at the sum frequency $\omega_{p}=\omega_{i}+\omega_{s}$ generated from the waves $\bm E_{s}(\bm r_{i})$, $\bm E_{i}(\bm r_{i})$ in the nonlinear structure can be presented as
\begin{equation}\label{eq:ESFG1}
E_{SFG,\sigma_{p}}(\bm r_{p})=\int \rmd ^{3}r_{0} G^{\vphantom{(2)}}_{\sigma_{p}\gamma}(\bm r_{p},\bm r_{0}) \chi^{(2)}_{\alpha\beta\gamma}(\bm r_{0})E_{s,\alpha}(\bm r_0)E_{i,\beta}(\bm r_0)\:.
\end{equation}
Inspired by linear reciprocity relationship Eq.~\eqref{eq:Lorentz}  we introduce the polarizations $P_{i,s,p}(\bm r)$ inducing the correspondent waves $E_{s,i,p}(\bm r)$,
\begin{equation}
\bm E_{\nu}(\bm r)=\int\rmd^{3}r_{0}\hat G(\bm r,\bm r_{0};\omega_{p})\bm P_{\nu}(\bm r_{0}),\quad \nu=i,s,p\:.
\end{equation}
This allows us to rewrite Eq.~\eqref{eq:Tsi1} and Eq.~\eqref{eq:ESFG1} as
\begin{equation}\label{eq:Tsi2}
\Psi(\bm r_{s},\bm r_{i},\sigma_{s},\sigma_{i},\omega_{s},\omega_{i})=\int\rmd^{3} r_{0}
\int\rmd^{3} r_{p}
G_{\sigma_{s}\alpha}(\bm r_{s},\bm r_{0})
G_{\sigma_{i}\beta}(\bm r_{i},\bm r_{0})\chi^{{(2)}}_{\alpha\beta\gamma}(\bm r_{0})G_{\gamma \sigma_{p}}(\bm r_{0},\bm r_{p})P_{p,\sigma_{p}}(\bm r_{p})\:,
\end{equation}
and 
\begin{equation}\label{eq:ESFG2}
E_{SFG,\sigma_{p}}(\bm r_{p};\omega_{i}+\omega_{s})=\int \rmd ^{3}r_{0}\int \rmd ^{3}r_{i}\int \rmd ^{3}r_{s} G^{\vphantom{(2)}}_{\sigma_{p}\gamma}(\bm r_{p},\bm r_{0}) \chi^{(2)}_{\alpha\beta\gamma}(\bm r_{0})G_{\alpha\sigma_{s}}(\bm r_0,\bm r_{s})G_{\alpha\sigma_{i}}(\bm r_0,\bm r_{i})
E_{s,\sigma_{s}}(\bm r_{s})E_{i,\sigma_{i}}(\bm r_{i})
\:.
\end{equation}
We have omitted the frequency arguments in the Green functions for the sake of brevity. In the reciprocal structure the Green functions satisfy the reciprocity property 
\begin{equation}\label{eq:Gab12}
G_{\alpha\beta}(\bm r_{1},\bm r_{2})=G_{\beta\alpha}(\bm r_{2},\bm r_{1})\:,
\end{equation}
that is equivalent to Eq.~\eqref{eq:Lorentz}. 
Comparing Eq.~\eqref{eq:Tsi2} and Eq.~\eqref{eq:ESFG2} with the help of Eq.~\eqref{eq:Gab12} we establish the general reciprocity relationship  between SPDC and SFG processes in the form
\begin{equation}\label{eq:gen-correspondence}
\iint \rmd^{3}r_{i}\rmd^{3}r_{s}
\Psi(\bm r_{s},\bm r_{i},\sigma_{s},\sigma_{i},\omega_{s},\omega_{i})P_{i,\sigma_{i}}(\bm r_{i})P_{s,\sigma_{s}}(\bm r_{i})=
\int \rmd^{3}r_{p}E_{SFG,\gamma}(\bm r_{p};\omega_{i}+\omega_{s})P_{p, \gamma}(\bm r_{p})\:.
\end{equation}

%%%%%%%%%%%%%%%%%%%%%%%%%%%%%%%%%%%%%%
\textbf{SPDC-SFG correspondence for a coupled waveguide array.}

In the previous section we have presented a general proof of the  SPDC-SFG correspondence for a reciprocal detection and excitation geometries. Here we apply this concept to the particular situation of   an array of coupled waveguides, parallel to the $z$ axis.
The waveguides can have arbitrary mode dispersion and losses and can be periodically patterned. 

{\it Green function expansion.~}  In what follows it will be useful to expand the Green function 
Eq.~\eqref{eq:Gtilde2}
over the set of the Bloch eigenmodes of the structure  with the Bloch wavenumbers $\beta_{n}$, characterizing propagation along $z$ direction at the given frequency $\omega$. The polarization degree of freedom is included in the index $n$ as well. Electric and magnetic fields of the eigenmodes  satisfy 
%the Maxwell equations
%\begin{equation}
%\rot \bm E_{n}=\frac{\rmi \omega}{c}\bm H_{n},\quad
%\rot \bm H_{n}=-\frac{\rmi \omega}{c}\eps(\bm r)\bm E_{n}\:
%\end{equation}
%and 
the orthogonality relation \cite{Sukhorukov:2014:OL,Chen:2010:PRA}
\begin{equation}
\iint \rmd x\rmd y [\bm E_{n'}\times\bm H_{n}-\bm E_{n}\times\bm H_{n'}]_{z}=s_{n}\delta_{n,-n'}\:,\label{eq:s}
\end{equation}
where $s_{n}$ is the so-called adjoint flux and we denote by the index $-n$ the solution with the wave vector $-\beta_{n}$. 
A pair of such forward and backward propagating solutions exists at any frequency provided that the structure is reciprocal.
The Green function  can be sought as an expansion over the eigenmodes propagating away from the point $z'$:
\begin{equation}
G_{\mu\nu}(\bm r,\bm r')=\begin{cases}
\sum\limits_{\Im \beta_{n}>0} u_{n,\nu}E_{n,\mu}(\bm r) & z>z'\\
\sum\limits_{\Im \beta_{n}<0} u_{n,\nu}E_{n,\nu}(\bm r) & z<z' \:, \label{eq:G0}
\end{cases}
\end{equation}
Here we distinguish between right- and left-propagating modes $\propto \e^{\pm \rmi \beta z}$ by the sign  of $\Im \beta$; for lossless medium an infinitely small losses can be formally  added to the permittivity. 
In order to find the expansion coefficients $u_{n,\beta}$  we use the Lorentz reciprocity theorem in the form \cite{Sukhorukov:2014:OL}
\begin{equation}
\frac{c}{4\pi}\oint_{S}\rmd S (\bm E\times\bm H_{n'}-\bm E_{n'}\times\bm H)=-\rmi \omega \int \rmd^{3}r\bm E_{n'}\cdot\bm P\:,
\label{eq:bio}
\end{equation}
where $\bm E$ and $\bm H$ are the electric and magnetic fields  induced by the dielectric polarization distribution $\bm P(\bm r)$. Namely, we replace $\bm E$ in Eq.~\eqref{eq:bio} by the Green function expansion 
Eq.~\eqref{eq:G0} and $\bm P(\bm r)$ by the point source term $\bm e_{\nu}\delta(\bm r-\bm r')$ where $\bm e_{\beta}$ is the unitary basis vector.  The integral in the left-hand side of Eq.~\eqref{eq:bio} is evaluated with the help of the  orthogonality relation Eq.~\eqref{eq:s}. This yields the equation for the coefficients 
$
s_{-n} u_{-n,\nu}=-4\pi \rmi \omega \bm e_{\nu}\cdot\bm E_{m}(\bm r')/c
$. Finding $u_{n,\nu}$ from this equation we present the Green function as 
\begin{equation}
G_{\alpha\beta}(\bm r,\bm r')=\sum\limits_{\Im \beta_{n}>0} f_n E_{n,\alpha}(\bm r_{>}) E_{-n,\beta}(\bm r_{<})\:, \label{eq:G1}
\end{equation}
where $f_n=-4\pi \rmi \omega/(c s_m)$ and $\bm r_{>}$ ($\bm r_{<}$) denotes one of the vectors $\bm r$, $
\bm r'$  with greater (lesser) coordinate $z$. 

{\it Proof of the SPDC-SFG correspondence.~}
The complex  wavefunction of a photon pair, generated in a $\chi^{(2)}$-nonlinear structure within the SPDC process, has the amplitude\cite{Poddubny:2016-123901:PRL}
\begin{equation}
T(\bm r_{s}\sigma_{s},\bm r_{i}\sigma_{i})=\int \rmd^{3} r_{0}
G_{\sigma_{s}\nu}(\bm r_{s},\bm r_{0},\omega_{s})G_{\sigma_{i}\mu}(\bm r_{i},\bm r_{0},\omega_{i})
 \chi^{(2)}_{\mu\nu;\eta}(\bm r_{0})E_{p,\eta}(\bm r_{0},\omega_{p})\:,\label{eq:T-SPDC}
\end{equation}
where $\bm r_{s} (\bm r_{i})$ and $\sigma_{s} (\sigma_{i})$ are signal (idler) photon coordinates and polarizations, respectively, and $E_{p}$ is the electric field of the pump with the frequency $\omega_{p}$. We now assume that the 
structure is pumped in the eigenmode $n_{p}$, substitute the Green function in the form Eq.~\eqref{eq:G1} and rewrite the biphoton wavefunction in the eigenmode representation as
\begin{equation}
\Psi(n_{p}\to n_{s},n_{i})=f_{n_{i}}f_{n_{s}}\int  \rmd^3 r_0  
\chi^{{(2)}}_{\mu\nu,\eta}(\bm r_0)E_{n_p,\eta}(\bm r_0)E_{-n_s,\mu}(\bm r_0)E_{-n_{i},\nu}(\bm r_0)\:.\label{eq:Psi_SPDC} 
\end{equation}
Now we consider the SFG process in the reverse direction. Two beams are injected into the ``signal'' and ``idler'' eigenmodes $-n_{i}$ and $-n_{s}$, propagating in the reverse direction. The generated SWM field is given by the convolution of the Green function with the nonlinear $\chi^{{(2)}}$-polarization induced by the incident waves,
\begin{equation}
E_{SFG}(\bm r)=\int\rmd^{3}r_{0}  G(\bm r,\bm r_{0})\chi^{{(2)}}_{\mu\nu,\eta}(\bm r_0)E_{-n_s,\mu}(\bm r_0)E_{-n_{i},\nu}(\bm r_0)\:.\label{eq:ESFG}
\end{equation}
Substituting the Green function expansion Eq.~\eqref{eq:G1} into Eq.~\eqref{eq:ESFG} we find the dimensionless SFG conversion amplitude   from the modes $-n_{s}$, $-n_{i}$ to the mode $-n_{p}$ propagating in the direction opposite to the pump of the SPDC process:
\begin{equation}
\xi(-n_{s},-n_{i}\to -n_{p})=f_{n_{p}}\int  \rmd^3 r_0  
\chi^{{(2)}}_{\mu\nu,\eta}(\bm r_0)E_{n_p,\eta}(\bm r_0)E_{-n_s,\mu}(\bm r_0)E_{-n_{i},\nu}(\bm r_0)\:.\label{eq:xiSFG}
\end{equation}
Comparing Eq.~\eqref{eq:xiSFG} and Eq.~\eqref{eq:Psi_SPDC} we establish our main result,  the correspondence between the biphoton wavefunction and the sum-frequency conversion efficiency
\begin{equation}
\Psi(n_{p}\to n_{s},n_{i})=\frac{f_{n_{i}}f_{n_{s}}}{f_{n_{p}}}\xi(-n_{s},-n_{i}\to -n_{p})\:.\label{eq:correspondence-main}
\end{equation}
We stress that the exact structure of the eigenmodes $\bm E_{m}$ was never used in the proof. The  only required property is the structure reciprocity in the linear regime, allowing to expand the Green function Eq.~\eqref{eq:G1} into the set of mutually reciprocal eigemodes $\bm E_{m}$ and $\bm E_{-m}$. As such, Eq.~\eqref{eq:correspondence-main} can be readily generalized to other geometries. For instance, if the Green function can expanded over the set of the solutions with   the asymptotic of outgoing and incoming spherical waves. Once the proper set of reciprocal solutions is determined, the SPDC-SFG correspondence relation can be established.

{\it Predicting the absolute SPDC photon count rate from the SFG conversion efficiency.} The SFG power conversion efficiency (erg/sec) can be found from the conversion amplitude 
Eq.~\eqref{eq:xiSFG} as 
\begin{equation}
\eta^{\rm SFG}_{n_s n_i} (\omega_s,\omega_i)=\frac{\phi_{n_p}}{\phi_{n_i}\phi_{n_s}}|\xi(-n_{s},-n_{i}\to -n_{p})|^2\:,
\label{eq:etaSFG}
\end{equation}
where $\phi_{n}=c\Re\iint \rmd x\rmd y \bm E_{n}\times\bm H^*_{n}/(2\pi)$ is the energy flux for the mode $n$.
In order to determine the photon coincidence count rate we need to calibrate the photon detection process~\cite{Poddubny:2016-123901:PRL}. To this end we explicitly introduce the signal and idler detectors modelled as the two-level systems with the dipole momenta matrix elements $\bm d_{i}$, $\bm d_{s}$ and the energies $\hbar \omega_{i}$, $\hbar \omega_{s}$. The number of photons absorbed by the detector per unit time 
is given by
\begin{equation}
\frac{\rmd N_{\rm abs}}{\rmd t}=\frac{2\pi}{\hbar}\delta(\hbar\omega-\hbar\omega_{i,s})|\bm d\cdot \bm E|^2,
\end{equation}
where $\bm E$ is the local electric field at the detector. On the other hand, the number of photons traveling through the given eigenmode $n$ per time is given by $\rmd N_{\rm phot}/\rmd t=\phi_{n}/\hbar\omega$. The ratio between the numbers of the absorbed photons and propagating photons provides the quantum efficiency of the detector for the mode $n$,
\begin{equation}
QE=\frac{N_{\rm abs}}{N_{\rm phot}}= \frac{2\pi \omega |\bm d\cdot \bm E|^2}{\hbar\phi_n}\:.
\end{equation}
The two-photon coincidence count rate per unit of the signal and idler spectra is formally defined as
\begin{equation}
\frac{\rmd N_{\rm pair}}{\rmd t \rmd \omega_i\rmd \omega_s}=\frac{W_{is}}{QE_iQE_s}
\label{eq:Npair}
\end{equation}
where
\begin{equation}
W_{is}=\frac{2\pi}{\hbar}\delta(\hbar\omega_p-\hbar\omega_i-\hbar\omega_s)|\sum \limits_{\sigma_{i}\sigma_{s}}d_{i}^{*}d_{s}^{*}T(\bm r_{s}\sigma_{s},\bm r_{i}\sigma_{i})|^2,
\end{equation}
is the uncalibrated rate of two photon counts calculated from the bi-photon amplitude Eq.~\eqref{eq:T-SPDC}.
In our geometry the generation and detection take place in the given eigenmodes $n_{s}$ and $n_{i}$. Substituting the definitions of the quantum efficiencies into Eq.~\eqref{eq:Npair} we obtain
\begin{equation}
\frac{\rmd N_{\rm pair}}{\rmd t \rmd \omega_i\rmd \omega_s}=
\frac{\delta(\omega_p-\omega_i-\omega_s)}{2\pi}\frac{\phi_{q_i}\phi_{q_s}}{ \omega_i\omega_s}
|\Psi(n_{p}\to n_{s},n_{i})|^{2}
 \frac{P_p}{\phi_{q_p}}\label{eq:Npair2}
\end{equation}
where $P_p ({\rm erg/sec})$ is the pump power (erg/sec). In order to 
compare the coincidence count rare  Eq.~\eqref{eq:Npair2}  with the sum frequency power conversion efficiency 
Eq.~\eqref{eq:etaSFG} we make use of our general SPDC-SFG link Eq.~\eqref{eq:correspondence-main}. 
This yields a general {\it absolute} correspondence between the sum frequency rate  and the photon pair generation rate:
\begin{equation}
\frac{1}{P_p}\frac{\rmd N_{\rm pair}}{\rmd t \rmd \omega_i\rmd \omega_s}=
\frac{\delta(\omega_p-\omega_i-\omega_s)}{2\pi }\frac{\phi_{q_i}^2\phi_{q_s}^2|f_{q_i}|^2|f_{q_s}|^2}{\omega_i\omega_s
\phi_{q_p}^2|f_{q_p}|^2}
\eta^{\rm SFG}_{n_s n_i} (\omega_s,\omega_i)\label{eq:absolute-correspondence}.
\end{equation}
\begin{figure*}
\centering
\includegraphics[width=\textwidth]{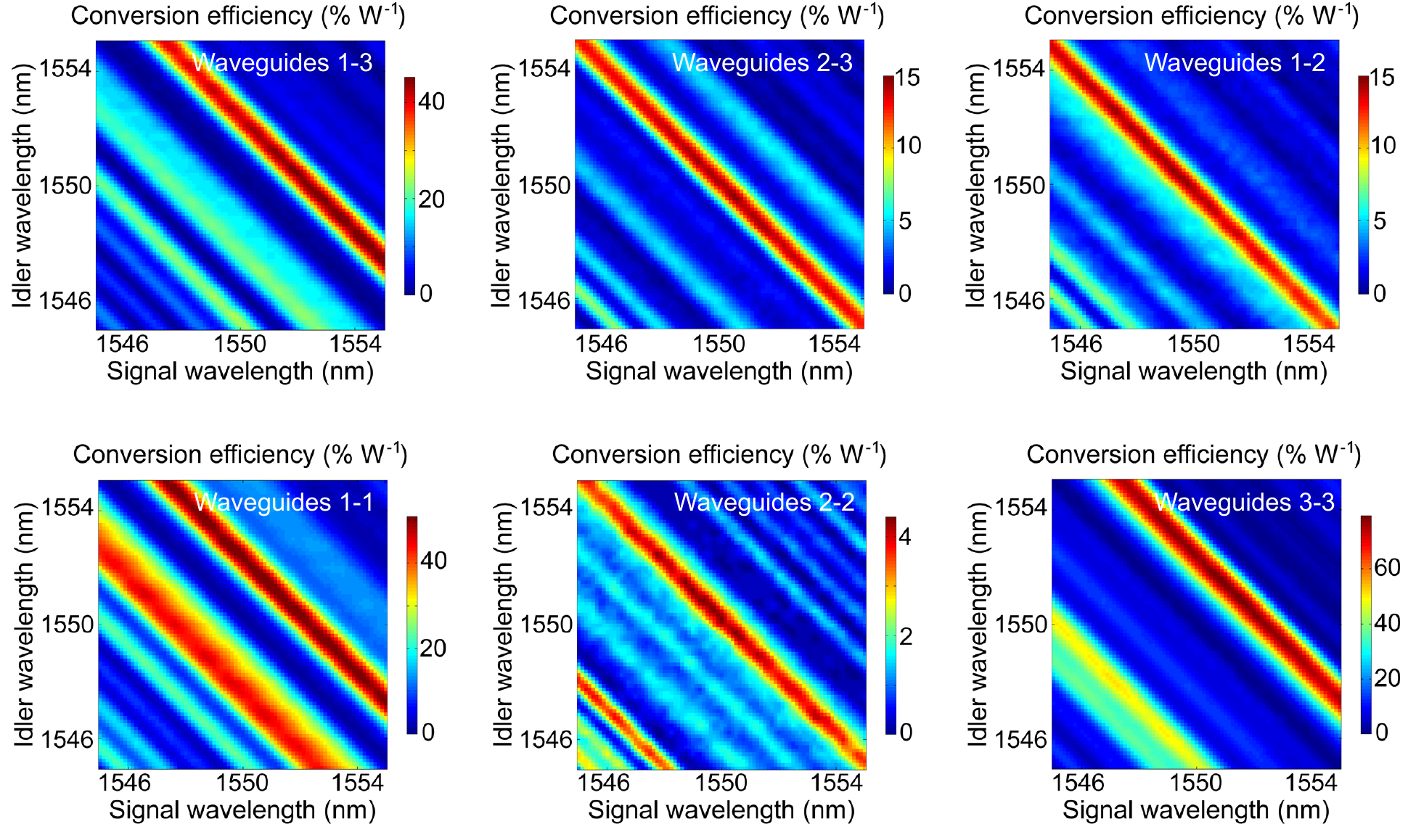}[t!]
\caption{\textbf{SFG measurements.} Measured classical sum-frequency conversion efficiency from waveguide 1 as a function of signal and idler wavelengths for all input combinations.
}
\label{fig1sp}
\end{figure*}
We now assume that in the detection region the tunneling coupling between the waveguides can be neglected and the flux and adjoint flux for each of the modes $n_{s}$, $n_{i}$, $n_{p}$ can be simplified to
\begin{equation}
\phi=\frac{c\mathcal {N} |E^2| S}{2\pi}, s=2S|E^2|\mathcal {N},\quad f=-\frac{2\pi \rmi \omega}{c S|E^2| \mathcal {N}}\:,
\end{equation}
where $S$ is a characteristic waveguide  area and $\mathcal {N}$ is the dimensionless mode refractive index. 
The products $\phi f$ in Eq.~\eqref{eq:absolute-correspondence} then reduce to $\omega_{i}$, $\omega_{s}$, $\omega_{p}$, respectively, and the correspondence law Eq.~\eqref{eq:absolute-correspondence}  assumes the simplified form

\begin{equation}
\frac{1}{P_p}\frac{\rmd N_{\rm pair}}{\rmd t \rmd \omega_i\rmd \omega_s}=
\frac{\delta(\omega_p-\omega_i-\omega_s)}{2\pi}\frac{\omega_i\omega_s}{\omega_p^2}\eta^{\rm SFG}_{n_s n_i} (\omega_s,\omega_i)\:.
\label{eq:absolute-correspondence2}
\end{equation}
Integrating Eq.~\eqref{eq:absolute-correspondence2} over the idler spectrum we recover Eq.~(2) in the main text linking the two-photon count rate per unit signal frequency to the sum frequency generation rate. 
\\

\textbf{SFG characterization data.}
Figure~\ref{fig1} shows the SFG efficiencies as a function of the wavelengths of signal and idler lasers for all input combinations. SFG power is measured from waveguide 1.
Corresponding measurements of photon pair generation via SPDC are shown in Fig.~\ref{fig2} for a 28.57 s acquisition time, a pump wavelength $\lambda_p=775 \ \mbox{nm}$, and power $P_p=32 \pm 5 \ \mu \mbox{W}$. Time bin width is 82~ps. Figure~\ref{fig3} shows the complete set of interferometric measurements used for phase reconstruction. The phases of the different elements $\Psi_{ij}$ were measured relative to $\Psi_{11}$.
%\newpage

\begin{figure*}
\centering
\includegraphics[width=\textwidth]{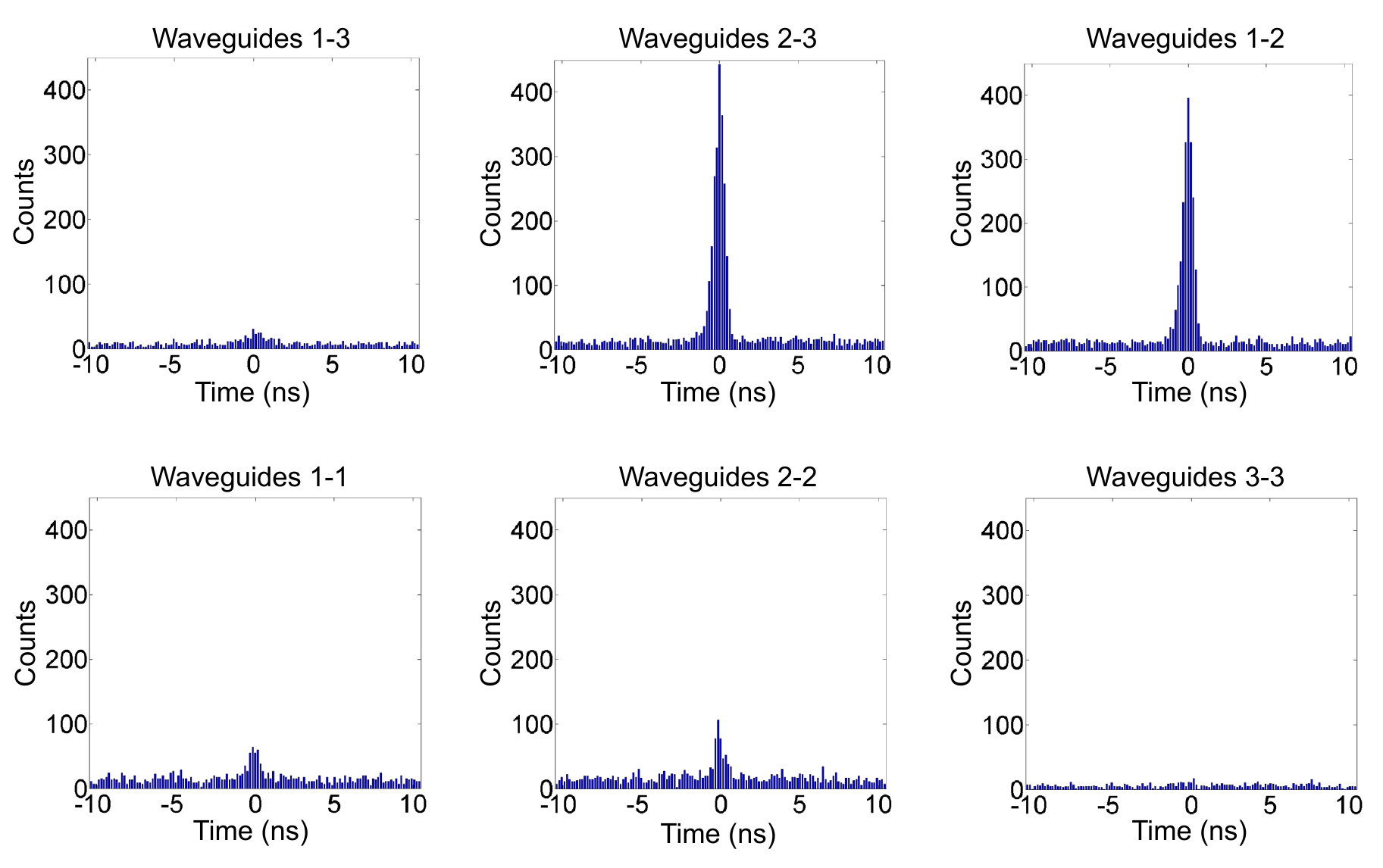}[h!]
\caption{\textbf{SPDC measurements.} Coincidence measurements for all output combinations when a pump beam with wavelength $\lambda_p=775 \ \mbox{nm}$, and power $P_p=32 \pm 5 \ \mu \mbox{W}$ is coupled to waveguide 1.}
\label{fig2sp}
\end{figure*}

\begin{figure*}
\centering
\includegraphics[width=\textwidth]{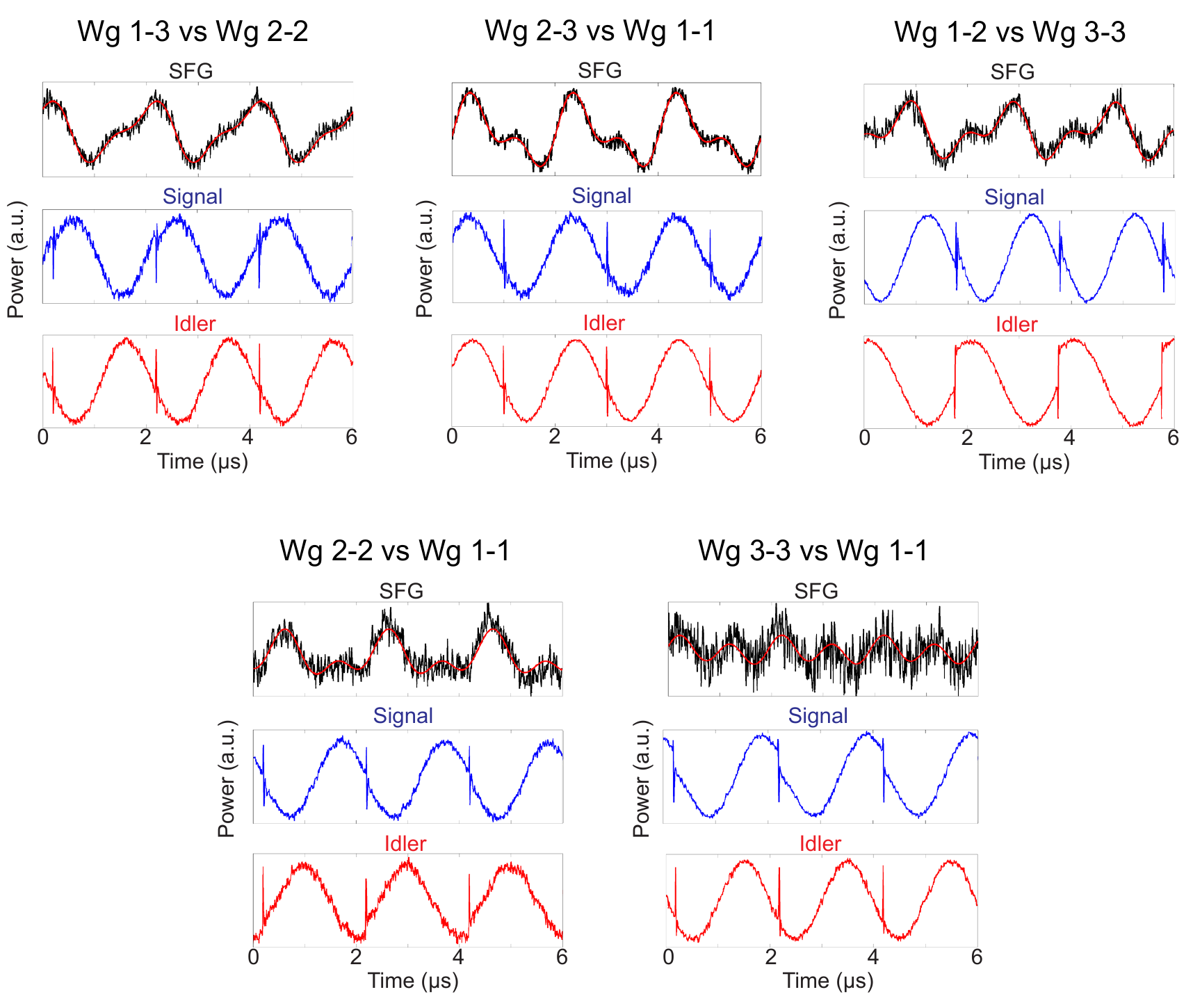}[h!]
\caption{\textbf{Phase measurements.} Oscilloscope traces obtained by collecting the beams with three different photodiodes for all input combinations. For each combination there are three traces that are used to measure the relative phase between the wave function elements $\Psi_{ij}$ and $\Psi_{11}$. Solid red line is the theoretical fit}
\label{fig3sp}
\end{figure*}
\end{widetext}
%\bibliography{db_sfg_paper_lenzini}
\end{document}